\documentclass[12pt]{article}

\usepackage[margin=1in]{geometry}

\usepackage{setspace}

\usepackage{graphicx}

\usepackage{enumitem}

\usepackage{microtype}
\usepackage{subfigure}
\usepackage{booktabs}
\usepackage{comment}
\usepackage{wrapfig}
\usepackage{arydshln}
\usepackage{enumitem}
\usepackage{multirow}
\usepackage{url}
\usepackage{natbib}
\usepackage{authblk}

\RequirePackage[colorlinks,citecolor=blue,linkcolor=blue,urlcolor=blue,pagebackref]{hyperref}

\usepackage{amsmath}
\usepackage{amssymb}
\usepackage{mathtools}
\usepackage{amsfonts}
\usepackage{bm}
\usepackage{bbm}

\usepackage{algorithm}
\usepackage{algorithmic}

\def\eqref#1{equation~\ref{#1}}

\def\1{\bm{1}}

\DeclareMathAlphabet{\mathsfit}{\encodingdefault}{\sfdefault}{m}{sl}
\SetMathAlphabet{\mathsfit}{bold}{\encodingdefault}{\sfdefault}{bx}{n}

\DeclareMathOperator*{\argmax}{arg\,max}

\newcommand{\p}[1]{\left(#1\right)}
\newcommand{\sqb}[1]{\left[#1\right]}
\newcommand{\cb}[1]{\left\{#1\right\}}

\usepackage{amsthm}

\theoremstyle{plain}

\usepackage[textsize=tiny]{todonotes}

\renewcommand{\eqref}[1]{(\ref{#1})}

\usepackage{xcolor}
\newcount\Comments  
\newcommand{\kibitz}[2]{\ifnum\Comments=1\textcolor{#1}{#2}\fi}

\usepackage[capitalize,noabbrev]{cleveref}

\allowdisplaybreaks

\title{Conformal Predictive Portfolio Selection}

\author{Masahiro Kato\thanks{Email: \texttt{mkato-csecon@g.ecc.u-tokyo.ac.jp}}$\,$}

\affil{Data Analytics Department, Mizuho-DL Financial Technology, Co., Ltd.}

\date{First version:  Oct 2024,  This version is of  \today}

\begin{document}

\maketitle

\begin{abstract}
This study examines portfolio selection using predictive models for portfolio returns. Portfolio selection is a fundamental task in finance, and a variety of methods have been developed to achieve this goal. For instance, the mean-variance approach constructs portfolios by balancing the trade-off between the mean and variance of asset returns, while the quantile-based approach optimizes portfolios by considering tail risk. These methods often depend on distributional information estimated from historical data using predictive models, each of which carries its own uncertainty. To address this, we propose a framework for portfolio selection via conformal prediction, called \emph{Conformal Predictive Portfolio Selection} (CPPS). Our approach forecasts future portfolio returns, computes the corresponding prediction intervals, and selects the portfolio of interest based on these intervals. The framework is flexible and can accommodate a wide range of predictive models, including autoregressive (AR) models, random forests, and neural networks. We demonstrate the effectiveness of the CPPS framework by applying it to AR models and neural networks and validate its performance through empirical studies, showing that it delivers superior returns compared to simpler strategies.
\end{abstract}

\section{Introduction}
Portfolio selection is a fundamental problem in finance, and numerous approaches have been developed to help investors construct desirable portfolios. A key aspect of building better portfolios is the use of estimated distributional information for future asset returns. In this study, given predictive models, including conventional autoregressive (AR) models and modern machine learning methods, we aim to develop a general framework for portfolio selection based on prediction intervals obtained through conformal prediction.

One of the primary approaches in portfolio selection is Markowitz's mean-variance portfolio theory, which optimizes portfolios by balancing the trade-off between the mean and variance of asset returns \citep{Markowitz1952,Markowitz1959,Markowitz2000}. Although widely adopted, the mean-variance approach has been criticized for relying on variance as a risk measure. Specifically, variance tends to increase with returns, even though higher returns are generally desirable. Moreover, variance considers the entire distribution of returns, including outcomes that might not reflect true risk from the investor's perspective. In response to these critiques, quantile-based approaches have gained traction. For example, \citet{Rockafellar2000OptimizationOC} propose minimizing Conditional Value at Risk (CVaR) through linear programming, while \citet{Bodnar2021} introduce a different quantile-based portfolio selection method that incorporates quantiles of both returns and risks.

Despite the development of various methods that leverage distributional information, a common challenge persists: relying on historical data alone may not yield accurate predictions. For instance, the historical sample mean can be a poor predictor of future asset returns. Because the ultimate goal is to optimize future returns, it may be beneficial to use predictive models, such as AR models or machine learning methods. Indeed, recent studies have employed machine learning approaches to forecast returns for diverse assets, including stocks, currencies, and real estate. However, both AR and machine learning models can complicate the assessment of prediction uncertainty. In traditional methods, such as low-dimensional linear regressions, confidence intervals are more straightforward to compute; in contrast, machine learning models typically involve high-dimensional parameters, making classical statistical inference more challenging. Additionally, under dependent data, it is difficult to obtain prediction intervals without imposing strong assumptions on the error term, such as normality.

This issue of uncertainty evaluation is especially relevant in finance. Conformal prediction addresses this concern by providing valid prediction intervals without requiring restrictive model assumptions \citep{Vovk2005,Chernozhukov2018Conformal}. Because of its model-free property, conformal prediction is an appealing tool for uncertainty evaluation in portfolio selection.

Building on this body of work, we propose a portfolio selection framework that employs prediction intervals. Our framework uses the confidence intervals of future asset returns generated by machine learning models and conformal prediction as the basis for its objective. In doing so, it provides a model-free, prediction-interval-based approach, allowing investors to define flexible portfolio objectives without pre-imposing a specific structure.

As an illustrative example, given a certain error level, our approach selects the portfolio with the highest predicted return within its confidence interval, ensuring that the lower bound of the return remains sufficiently high under the chosen error threshold. In this process, we forecast future returns for each portfolio candidate, construct prediction intervals using conformal prediction, and then select portfolios based on their predicted returns at a specified error rate. This strategy aims to improve the worst-case performance of the selected portfolio.

Important related work includes research on portfolio selection within a Bayesian framework, which provides a way to measure the uncertainty of future asset returns \citep{Barry1974,brown1976optimal,Winkler1975}. The Bayesian approach has been applied to mean-variance portfolios by \citet{Bauder2021} and to quantile-based portfolios by \citet{bodnar2020bayesian}. More recent studies, such as \citet{tallman2023bayesian} and \citet{kato2024bpps,kato2024generalbayesianpredictivesynthesis}, explore Bayesian ensemble methods for portfolio selection.

In our algorithm, we propose applying conformal prediction for each portfolio return rather than for individual asset returns, though our method require conducts conformal prediction as the number of portfolio candidates. This is because conformal prediction has traditionally relied on univariate methods, making it challenging to directly handle correlations among multiple assets or jointly predict entire asset-return vectors. Until late 2024, multivariate conformal prediction methods were not well-established, leading some approaches—including ours—to avoid full multivariate modeling by computing intervals for aggregated portfolios rather than individual assets. However, recent work (early 2025) by \citet{thurin2025optimaltransportbasedconformalprediction} and \citet{klein2025multivariateconformalpredictionusing} employs optimal transport to achieve multivariate conformal prediction, suggesting the possibility of handling asset-return vectors without resorting to portfolio aggregation. While these methods, if adapted to time-series data, could offer more comprehensive coverage of multivariate dynamics, our univariate-based approach remains simpler to implement. It avoids the complexities of full multivariate modeling—even if it demands heavier computation when many portfolios are considered—and can therefore still be advantageous in practical settings.

\section{Problem Setting}
Let $T, K \geq 2$ be positive integers. Consider a time series with $T+1$ periods denoted by $1,2,\dots,T, T+1$. There are $K$ financial assets, and each asset $a \in [K] \coloneqq \{1, 2, \dots, K\}$ yields a return $Y_{a, t}$ in each period $t \in [T+1]$. Additionally, for each period $t \in [T+1]$, there is a $d$-dimensional feature vector $X_{a, t} \in \mathcal{X} \subseteq \mathbb{R}^d$, where $\mathcal{X}$ is a space of feature vectors. These feature vectors are used to predict future asset returns or portfolio returns. The vector $X_{a, t}$ can incorporate both endogenously generated variables and historical target variables observed up to period $t-1$, such as $Y_{a, 1}, Y_{a, 2}, \dots, Y_{a, t-1}$, but it cannot include the target variable $Y_{a, t}$ observed in period $t$. We denote the sets of returns and features for the $K$ assets by $Y_t = (Y_{a, t})_{a\in[K]}$ and $X_t = (X_{a, t})_{a\in[K]}$, respectively.

We refer to a ratio $\bm{w} \in \widetilde{\mathcal{W}}$ of the investment as the portfolio, where
\[
\widetilde{\mathcal{W}} \coloneqq \{\bm{w} \coloneqq \{w_{a, T+1}\}_{a \in [K]} \in [0, 1]^K \mid \sum_{a \in [K]} w_{a, T+1} = 1\}.
\]
By holding a portfolio $\bm{w}$, we obtain a return 
\[
R_{T+1}(\bm{w}) \coloneqq \sum^K_{a=1}w_{a, T+1} Y_{a, T+1}
\]
after the portfolio is selected.

We assume the dataset $\{(Y_t, X_t)\}_{t=1}^T$ and the feature vector $X_{T+1}$ are observable at period $T+1$, and we can use them to select a portfolio. In this study, our task is to select a desirable portfolio in period $T+1$, given $\mathcal{W} \subset \widetilde{\mathcal{W}}$, where $\mathcal{W}$ is a finite subset of $\widetilde{\mathcal{W}}$. In period $T+1$, based on this dataset and $X_{T+1}$, we select a portfolio 
\[
\bm{w}_{T+1} \in \mathcal{W},
\]
which then yields a return 
\[
R_{T+1}(\bm{w}_{T+1}) \coloneqq \sum^K_{a=1}w_{a, T+1} Y_{a, T+1}
\]
after the portfolio is formed. Note that $(Y_{a, T+1})_{a\in[K]}$ is unobservable before constructing the portfolio, whereas the feature vector $X_{T+1}$ is observable. Our objective is to select a portfolio that satisfies an investor's criterion.

In portfolio selection, investors typically account for both the uncertainty of asset returns and their individual risk preferences. Simply maximizing $R_{T+1}(\bm{w}_{T+1})$ may not be desirable because such an approach can imply taking on excessive risk. To balance the trade-off between returns and risk, many portfolio objectives incorporate risk measures, such as variances or quantiles. Notable examples include the mean-variance portfolio, the risk-parity portfolio, and various quantile-based strategies.

These existing approaches depend on distributional information (e.g., means, variances, and quantiles), which is unknown and must be estimated for portfolio construction. This estimation introduces uncertainty due to estimation errors and the potential for distributional shifts over time. In particular, because our focus is on the distributional information of future returns, a range of predictive models is commonly employed, including modern machine learning algorithms. Yet, these methods may complicate uncertainty quantification, given their complexity relative to more classical models such as linear regression.

Here, we propose a technique for constructing portfolios using prediction intervals that more effectively capture the uncertainty in estimators of distributional information.

\subsection{Predictive Models}
Our focus is on portfolio selection in period $T+1$, given $\{(Y_t, X_t)\}_{t=1}^T$ and $X_{T+1}$. Since the portfolio return $R_{T+1}(\bm{w}_{T+1})$ is a future, unobserved value, we use various predictive models to forecast it.

We formalize the setting as follows. Given $\{(Y_t, X_t)\}_{t=1}^T$ and $X_{T+1}$, for each $\bm{w} \in \mathcal{W}$, we predict the portfolio return $R_{T+1}(\bm{w}_{T+1})$ using models such as linear regression, random forests, and neural networks. These predictive models can be trained or estimated using the dataset $\{(Y_t, X_t)\}_{t=1}^T$ and $X_{T+1}$. In time series analysis, standard methods include AR models and moving-average (MA) models \citep{Hamilton1994}. 

\subsection{Conformal prediction  of Portfolio Return}
We construct portfolios based on the predictions generated by these models. To measure the uncertainty of these predictions, we employ conformal prediction, which is flexible because it does not impose specific restrictions on the choice of predictive models, aside from certain conditions such as estimation error rates.

Let $\alpha \in (0, 1)$ be an error rate. Using conformal prediction, given the dataset $\{(Y_t, X_t)\}_{t=1}^T$ and a portfolio $\bm{w} \in \mathcal{W}$, we construct a prediction interval $\widehat{C}^{\bm{w}}_T(X_{T+1})$ satisfying
\begin{align*}
    \mathbb{P}\p{ R_{T+1}(\bm{w}) \in \widehat{C}^{\bm{w}}_T(X_{T+1}) } \geq 1 - \alpha,
\end{align*}
where the probability $\mathbb{P}$ is taken over $\{(Y_t, X_t)\}_{t=1}^{T+1}$.

\section{Conformal Predictive Portfolio Selection}
This study employs prediction intervals for future asset returns to guide portfolio selection. While predictive asset returns offer insights into prospective performance, they often fail to convey the associated uncertainty. In portfolio selection, particularly when investors are not risk-neutral, this uncertainty strongly influences the choice of a desirable portfolio. Hence, it is crucial to include a method that accounts for the uncertainty in forecasted portfolio returns.

To this end, we use conformal prediction, which formally quantifies the uncertainty of predictions. Conformal prediction provides a prediction interval 
\[
\widehat{C}^{\bm{w}}_T(X_{T+1})
\] 
such that 
\[
\mathbb{P}\p{R_t\p{\bm{w}} \in \widehat{C}^{\bm{w}}_T(X_{T+1})} = 1 - \alpha.
\] 
In this study, for each \(\bm{w} \in \mathcal{W}\), we compute the prediction interval \(\widehat{C}^{\bm{w}}_T(X_{T+1})\) using conformal prediction, not for each asset return, and optimize an objective that depends on these intervals.

We define the mechanism that takes prediction intervals as input and returns a portfolio \(\widehat{\bm{w}}_{T+1}\) by
\[
    \mathrm{PI}\p{\cb{\widehat{C}^{\bm{w}}_T(X_{T+1})}_{\bm{w}\in\mathcal{W}}} = \widehat{\bm{w}}_{T+1}.
\]
A portfolio obtained in this manner is called a \emph{prediction-interval (PI)-based portfolio}.

Our framework is flexible and can accommodate a range of objectives for portfolio selection, allowing freedom in both the choice of predictive models and conformal prediction  methods. We do not impose specific choices for these components, as suitable methods may differ according to the data-generating process. For instance, for dependent data, one could use the conformal prediction  techniques proposed by \citet{Chernozhukov2018Conformal}; the appropriate methods should be selected based on the nature of the data.

We refer to our framework as \emph{conformal predictive portfolio selection} (CPPS), which uses conformal prediction  to generate prediction intervals and then exploits those intervals to construct PI-based portfolios. Our CPPS method has two main steps:
\begin{itemize}
    \item[\(\bullet\)] For each portfolio \(\bm{w} \in \mathcal{W}\), compute a prediction interval \(\widehat{C}^{\bm{w}}_T(X_{T+1})\) using conformal prediction, and evaluate the portfolio value.
    \item[\(\bullet\)] Select the desirable portfolio by choosing the one that achieves the best value based on the prediction intervals.
\end{itemize}
Pseudo-code for this procedure is given in Algorithm~\ref{alg:cpps}.

The reason we opt to calculate predictive intervals on a portfolio-by-portfolio basis, rather than for each individual asset $Y_{a, T+1}$ relates to the need to capture the correlation structure across multiple assets. In principle, one would want a multivariate conformal prediction framework to handle such correlations comprehensively. However, as of October 2024 (when this paper was released), multivariate conformal prediction methods remained insufficiently developed. For this problem, our approach proposes merging the returns of multiple assets into a single portfolio return in advance of the conformal prediction, thus bypassing the complexities associated with multivariate time-series analysis. The trade-off, however, is that this approach scales with the number of portfolios: it can only be run on a finite set of candidates and can become computationally intensive as the number of portfolios increases.

In early 2025, two papers, \citet{thurin2025optimaltransportbasedconformalprediction} and \citet{klein2025multivariateconformalpredictionusing}, presented multivariate conformal prediction methods using optimal transport. If extended to time-series data, these methods could yield direct prediction intervals on the asset-return vector, eliminating the need to bundle returns into portfolios purely to avoid multivariate challenges. We view such methods as a promising direction for future work, potentially enabling a conformal-interval construction for fully multivariate financial time series.

Nevertheless, our current approach remains appealing for two key reasons. First, modeling high-dimensional time series directly is often difficult, whereas our procedure only requires univariate forecasting models. Second, our proposed method is simple in implementation: we rely on a standard conformal prediction routine and a relatively straightforward algorithmic structure. By contrast, multivariate conformal procedures based on optimal transport involve more intricate algorithms. This simplicity plays an important role in practice. Hence, although our approach may be more demanding computationally when many portfolios are considered, it retains practical and implementational advantages in real-world scenarios.

In practice, one may limit the portfolio class \(\mathcal{W}\) to a finite set. Addressing the reduction of computational costs remains an important direction for future research.

\begin{algorithm}[tb]
   \caption{CPPS}
   \label{alg:cpps}
\begin{algorithmic}
   \STATE {\bfseries Input:} Predictive models, portfolio candidates $\mathcal{W}$, and error rate $\alpha \in (0, 1)$.
   \FOR{$\bm{w}\in\mathcal{W}$}
   \STATE Conduct conformal prediction  for the predictive models and obtain the prediction interval $\widehat{C}^{\bm{w}}_T(X_{T+1})$ for predicting $R_{T+1}(\bm{w})$.
   \ENDFOR
   \STATE Obtain $\widehat{\bm{w}}_{T+1} = \mathrm{PI}\p{\{\widehat{C}^{\bm{w}}_T(X_{T+1})\}_{\bm{w}\in\mathcal{W}}}$. 
\end{algorithmic}
\end{algorithm}

\begin{algorithm}[tb]
   \caption{HR--LR CPPS}
   \label{alg2}
\begin{algorithmic}
   \STATE {\bfseries Input:} Predictive models, portfolio candidates $\mathcal{W}$, error rate $\alpha \in (0, 1)$, and $m\in\mathbb{N}$
   \FOR{$\bm{w}\in\mathcal{W}$}
   \STATE Conduct conformal prediction  and obtain $\widehat{C}^{\bm{w}}_T(X_{T+1})$. 
   \STATE Define $\overline{r}^{\bm{w}, \alpha}_{T+1}$ and $\underline{r}^{\bm{w}, \alpha}_{T+1}$.
   \ENDFOR
   \STATE Choose $m$ portfolios from the lowest returns to the $m$-th lowest returns and denote the resulting portfolios by $\underline{\mathcal{W}}$.
   \STATE Select $\bm{w} = \argmax_{\bm{w}\in\underline{\mathcal{W}}}\overline{r}^{\bm{w}, \alpha}_{T+1}$.
\end{algorithmic}
\end{algorithm}

\subsection{Example: HR--LR CPPS}
Although our CPPS framework does not mandate a particular choice of predictive models or conformal prediction  methods, it is illustrative to present a concrete example. Here, we provide an example of CPPS in which we select a portfolio aimed at maximizing returns at a specified error rate \(\alpha\), while constraining risk. We call this the High-Return-from-Low-Risk (HR--LR) portfolio. Although simple, this example clarifies how to apply the CPPS framework. We also offer the full procedure and the corresponding pseudo-code. For conformal prediction, we adopt the method proposed by \citet{Chernozhukov2018Conformal}.

Let \(\alpha \in (0, 1)\) be the error rate, and let \(\mathcal{H}\) be the hypothetical values of \(R_{T+1}(\bm{w})\). For simplicity, assume that \(\mathcal{H}\) is a discrete set, for example \(\mathcal{H} = \{-0.3, -0.2, 0.0, 0.1, 0.2, 0.3\}\). a

For each \(\bm{w} \in \mathcal{W}\), we use conformal prediction  to produce a prediction interval
\[
\widehat{C}^{\bm{w}}_T(X_{T+1}) \subseteq \mathcal{H}
\]
such that
\[
\mathbb{P}\p{R_{T+1}(\bm{w}) \in \widehat{C}^{\bm{w}}_T(X_{T+1})} \ge 1 - \alpha.
\]

Let \(m \ge 1\) be a positive integer. For each portfolio \(\bm{w} \in \mathcal{W}\), denote the lowest and highest returns in \(\widehat{C}^{\bm{w}}_T(X_{T+1})\) by \(\underline{r}^{\bm{w}, \alpha}_{T+1}\) and \(\overline{r}^{\bm{w}, \alpha}_{T+1}\), respectively. We select \(m\) portfolios from \(\mathcal{W}\) whose \(\underline{r}^{\bm{w}, \alpha}_{T+1}\) values are among the top \(m\) in terms of the lower bound. Denote this new set by \(\underline{\mathcal{W}} \subset \mathcal{W}\). We then pick the portfolio from \(\underline{\mathcal{W}}\) whose \(\overline{r}^{\bm{w}, \alpha}_{T+1}\) is highest:
\[
\bm{w}^{\mathrm{HR}\mathchar`-\mathrm{LR}} 
= \argmax_{\bm{w} \in \underline{\mathcal{W}}} \overline{r}^{\bm{w}, \alpha}_{T+1}.
\]
This portfolio is expected to exhibit the highest potential return among the portfolios whose predicted lower return bound is relatively large.

\subsection{HR--LR CPPS with AR Models}
We now give a more concrete demonstration of the CPPS framework using AR models as the predictive tool. For conformal prediction, we again use the method of \citet{Chernozhukov2018Conformal}.

\subsubsection*{Step~1: Data Augmentation}
Let hypothetical values \(\mathcal{H} = \{r^{(1)}, r^{(2)}, \dots, r^{(H)}\}\) be given. For each \(\bm{w} \in \mathcal{W}\) and \(r \in \mathcal{H}\), define an augmented dataset \(\mathcal{D}_{(r)} = \{Z_t\}_{t=1}^{T+1}\), where
\begin{align}
    \label{eq:dataset}
    Z_t = \p{\widetilde{R}_t, X_t} = 
\begin{cases}
\p{R_t(\bm{w}), X_t} & \text{if } 1 \le t \le T,\\
(r, X_t) & \text{if } t = T+1.
\end{cases}.
\end{align}

Let \(\pi\) be a permutation of \(\{1, 2, \dots, T\}\). Denote the permuted dataset by \(\mathcal{D}^\pi_{(r)} = \{Z_{\pi(t)}\}_{t=1}^T\). We assume that the identity permutation \(\mathbb{I}\) belongs to the set of permutations, so that \(\mathcal{D}_{(r)} = \mathcal{D}^{\mathbb{I}}_{(r)}\). Following \citet{Chernozhukov2018Conformal}, we specifically use a blocking permutation, defined by
\[
    t \mapsto \pi_j(t) = 
    \begin{cases}
        t + (j - 1) & \mathrm{if}\ 1 \le t \le T - (j-1),\\
        t + (j - 1) - T & \mathrm{if}\ T - (j-1) + 1 \le t \le T,
    \end{cases}
\]
for \(t = 1,\dots, T\).

\subsubsection*{Step~2: Training a Predictive Model}
For each dataset 
\[\mathcal{D}^\pi_{(r)} = \cb{\p{\widetilde{R}_{\pi(t)}, X_{\pi(t)}}}_{t=1}^{T+1},\] 
including the original data \(\mathcal{D}_{(r)}\), we train an AR model on \(\cb{\p{\widetilde{R}_{\pi(t)}, X_{\pi(t)}}}_{t=1}^T\). Denote the trained model by \(f^\pi_T\), with \(f_T\) corresponding to the model trained on \(\mathcal{D}_{(r)}\) (i.e., the identity permutation).

\subsubsection*{Step~3: Conformal prediction }
We define the $p$-value as
\begin{align}
\label{eq:conform_p}
    \widehat{p}(r) := \frac{1}{|\Pi|} \sum_{\pi \in \Pi} \mathbbm{1}[ S(\mathcal{D}^\pi_{(r)}) \geq S(\mathcal{D}_{(r)}) ],
\end{align}
where $S(\cdot)$ is the nonconformity score. In this case, $S(\cdot)$ is defined as the (empirical) mean squared error between the predicted values and $\widetilde{R}_{\pi(t)}$:
\begin{align}
    \label{eq:score}
    &S(\mathcal{D}_{(r)}) = \frac{1}{T+1}\sum^{T+1}_{t=1}\left(\widetilde{R}_{t} - f_{T}(X_{t})\right)^2,\\
    &S(\mathcal{D}^\pi_{(r)}) = \frac{1}{T+1}\sum^{T+1}_{t=1}\left(\widetilde{R}_{\pi(t)} - f^\pi_{T}(X_{\pi(t)})\right)^2.\nonumber
\end{align}
For an error rate \(\alpha \in (0, 1)\), the prediction set is defined as
\[
\widehat{C}^{\bm{w}}_T(X_{T+1}) = \cb{r : \widehat{p}(r) > \alpha }.
\]
We evaluate this on a grid \(\mathcal{H}\). Pseudo-code for this conformal prediction  procedure, based on \citet{Chernozhukov2018Conformal}, is presented in Algorithm~\ref{alg:CP}.

\subsubsection*{Step~4: Defining Highest Return and Lowest Risk}
For each \(\bm{w}\), define 
\(\overline{r}^{\bm{w}, \alpha}_{T+1} = \max_{r \in \widehat{C}^{\bm{w}}_T(X_{T+1})} r\) 
and 
\(\underline{r}^{\bm{w}, \alpha}_{T+1} = \min_{r \in \widehat{C}^{\bm{w}}_T(X_{T+1})} r.\)
We then pick the \(m\) portfolios whose \(\underline{r}^{\bm{w}, \alpha}_{T+1}\) values are highest and collect them into \(\underline{\mathcal{W}} \subset \mathcal{W}\). 

\subsubsection*{Step~5: HR--LR CPPS}
Finally, from the set \(\underline{\mathcal{W}}\), we select the portfolio with the largest \(\overline{r}^{\bm{w}, \alpha}_{T+1}\):
\[
\widehat{\bm{w}}_{T+1} 
= \argmax_{\bm{w} \in \underline{\mathcal{W}}} \overline{r}^{\bm{w}, \alpha}_{T+1}.
\]

Thus, in the HR--LR procedure, we first look for 
portfolios whose \emph{lower bound} of the $(1-\alpha)$-prediction interval is comparatively large, thus controlling downside risk. Concretely, for each candidate portfolio $\bm{w}$, let 
$\underline{r}^{\bm{w}, \alpha}_{T+1} = \min\,\widehat{C}^{\bm{w}}_T(X_{T+1})$ 
denote the minimal return in the conformal prediction set. A high 
$\underline{r}^{\bm{w}, \alpha}_{T+1}$ suggests that, with high probability, 
the portfolio's return will not fall below that threshold. 
From among those ``low-risk'' portfolios (i.e., those with high lower bounds), 
we then pick the one with the largest \emph{upper bound} 
$\overline{r}^{\bm{w}, \alpha}_{T+1} = \max\,\widehat{C}^{\bm{w}}_T(X_{T+1})$, 
thereby aiming for high potential upside. This two-stage selection encapsulates 
an intuitive trade-off: select a portfolio with strong worst-case protection 
\emph{and} attractive best-case performance.

\begin{algorithm}
  \caption{Conformal prediction }\label{alg:CP}
  \begin{algorithmic}
   \STATE {\bfseries Input:} Data $\{(X_t, Y_t)\}^T_{t=1}$, $X_{T+1}$, portfolio $\bm{w}$, error rate $\alpha \in (0,1)$, and hypothesis values $\mathcal{H}$. 
   \FOR{$r \in \mathcal{H} \subset \mathbb{R}^{T_1}$}
   \STATE Define $Z_{(y)}$ as in \eqref{eq:dataset}.
   \STATE Compute $\widehat{p}(r)$ using \eqref{eq:conform_p}.
   \ENDFOR
   \STATE {\bfseries Return:} The $(1-\alpha)$ confidence interval $\widehat{C}^{\bm{w}}_T(X_{T+1}) = \{r : \widehat{p}(r) > \alpha\}$.
\end{algorithmic}
\end{algorithm}

\subsection{Theoretical Analysis}
We now justify the application of conformal prediction  for dependent data, following the framework in \citet{Chernozhukov2018Conformal}.

Let \(S_*\) be an unobserved oracle score function. The validity of conformal prediction  in a dependent-data setting depends on how accurately the score \(S\), defined in \eqref{eq:score}, approximates \(S_*\).

When AR models are used with the blocking permutation \(\Pi\), and under certain regularity conditions, the following statements hold for sequences \(\{\delta_{1,t}, \delta_{2,t}, \gamma_{1,t}, \gamma_{2,t}\}_{t=1}^T\), where each term converges to zero as \(t \to \infty\) \citep{Chernozhukov2018Conformal}:
\begin{itemize}
    \item With probability at least \(1 - \gamma_{1}\), the randomization distribution
    \[
    \widetilde{F}(x) \coloneqq \frac{1}{T}\sum_{\pi\in\Pi}\mathbbm{1}\sqb{S_*\p{Z^\pi} < x}
    \]
    satisfies
    \[
    \bigl|\widetilde{F}(x) - F(x)\bigr| \le \delta_{1, T},
    \]
    where \(F(x) = P\p{S_*(Z) < x}\). When this holds, we say that \(\widetilde{F}(x)\) is approximately ergodic for \(F(x)\).
    \item With probability at least \(1 - \gamma_{2}\), the estimation errors are small:
    \begin{itemize}
        \item The mean squared error satisfies 
        \[
        \frac{1}{T}\sum_{\pi \in \Pi}\p{S\p{Z^\pi} - S_*\p{Z^\pi}}^2 \le \delta^2_{2, T}.
        \]
        \item The pointwise error at \(\pi = \mathrm{Identity}\) is small: 
        \[
        \bigl|S(Z) - S_*(Z)\bigr| \le \delta_{2, T}.
        \]
        \item The density of \(S_*(Z)\) is uniformly bounded by a constant \(D\).
    \end{itemize}
\end{itemize}
Note that the number of permutations satisfies \(|\Pi| = T\).

Therefore, the confidence interval derived from conformal prediction  has approximate coverage of \(1-\alpha\). Formally,
\[
\begin{aligned}
&\bigl|\mathbb{P}\p{R_{T+1}(\bm{w}) \in \widehat{C}^{\bm{w}}_T(X_{T+1})} - (1 - \alpha)\bigr|\\
&\qquad\le 6\delta_{1, T} + 4\delta_{2, T} + 2D\p{\delta_{2, T} + 2\sqrt{\delta_{2, T}}} + \gamma_{1, T} + \gamma_{2, T}.
\end{aligned}
\]
This result indicates that under our framework, the confidence interval obtained from conformal prediction  is approximately valid, justifying the HR--LR CPPS approach with AR models.

\section{Experiments}
In this section, we explore the empirical performance of our proposed CPPS framework, focusing on the HR--LR CPPS. We conduct empirical studies using stock data from the US and Japanese markets. Specifically, the HR--LR CPPS employs AR($3$) models and neural networks to construct predictive intervals. In the neural network approach, we use a feedforward network with $100$ hidden units, taking as input $R_{T}(\bm{w}), R_{T-1}(\bm{w}), R_{T-3}(\bm{w})$ to predict $R_{T+1}(\bm{w})$ for each $\bm{w} \in \mathcal{W}$. In each market, we select three representative stocks, as listed in Tables~\ref{tab:us} and \ref{tab:jap}.

The stock price data covers the period from January 1, 2009, to December 31, 2018. Returns are calculated monthly. Data from January 2009 through December 2011 is used exclusively for parameter learning, and portfolio performance is tested on data from January 2012 to December 2018. We sequentially update the parameter estimates after January 2012. 

\begin{table}[t]
    \caption{US stock data}
    \label{tab:us}
    \centering
    \scalebox{1.0}{
    \begin{tabular}{|c|c|}
    \hline
    Company & Industry \\
    \hline
    Apple Inc. & Technology \\ 
    Microsoft Corp. & Technology \\ 
    Amazon.com Inc. & Consumer Discretionary \\ 
    \hline
    \end{tabular}
    }
\end{table}

\begin{table}[t]
    \caption{Japanese stock data}
    \label{tab:jap}
    \centering
    \scalebox{1.0}{
    \begin{tabular}{|c|c|}
    \hline
    Company & Industry \\
    \hline
    Toyota Motor & Automotive \\
    SoftBank Group & Telecommunication \& IT \\
    Keyence & Electronic Equipment \\ 
    \hline
    \end{tabular}
    }
\end{table}

\begin{figure}[H]\centering
\includegraphics[width=0.8\linewidth]{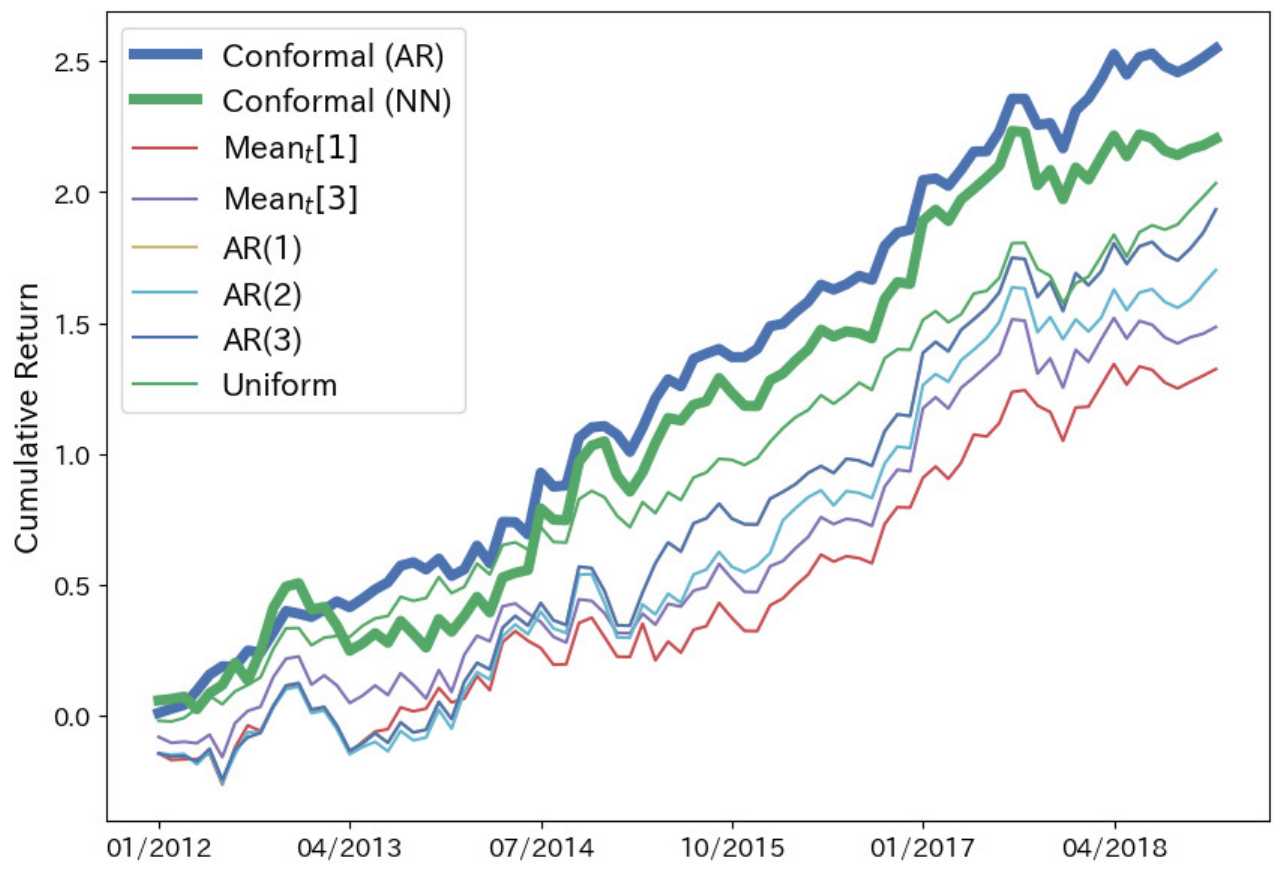}
\caption{Experimental results for US stocks. The $y$-axis indicates cumulative returns, and the $x$-axis shows months and years.}
\label{fig:fig1}

\includegraphics[width=0.8\linewidth]{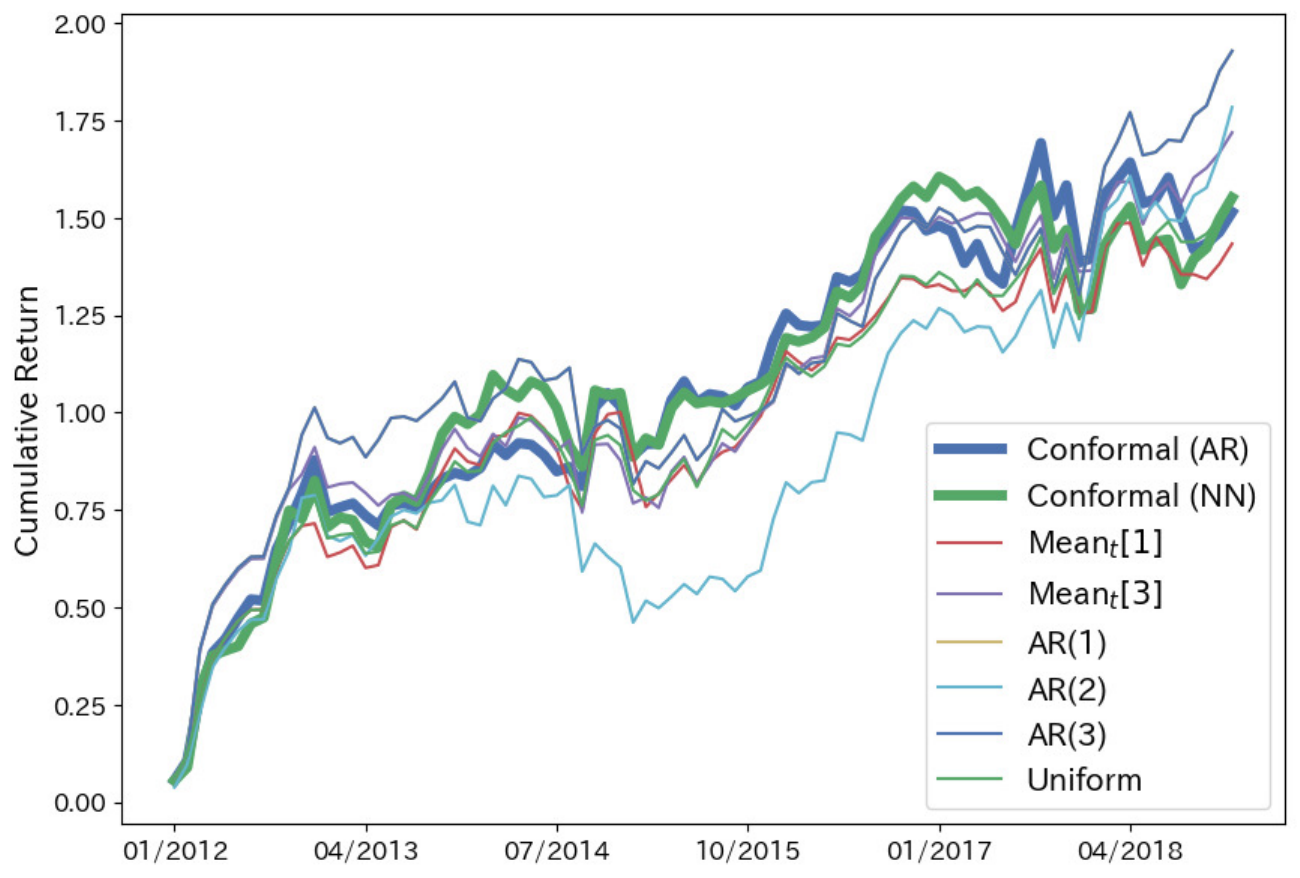}
\caption{Experimental results for Japanese stocks. The $y$-axis indicates cumulative returns, and the $x$-axis shows months and years.}
\label{fig:fig2}
\end{figure}

\subsection{Alternative Methods}
For comparison, we construct portfolios using the following approaches:
\begin{itemize}
    \item The sample mean over the past 1 year ($Mean_t[1]$).
    \item The sample mean over the past 3 years ($Mean_t[3]$).
    \item An AR$(1)$ regression model using samples from the past 3 years ($AR(1)$).
    \item An AR$(2)$ regression model using samples from the past 3 years ($AR(2)$).
    \item An AR$(3)$ regression model using samples from the past 3 years ($AR(3)$).
    \item An equal-investment strategy that allocates $1/K$ to each asset (Uniform).
\end{itemize}

\subsection{Experimental Results}
We apply each method to the dataset spanning January~1, 2008, through December~31, 2019, and report their cumulative returns.\footnote{We assume that investors can adjust their portfolio holdings without incurring additional transaction costs.} 

Figures~\ref{fig:fig1} and \ref{fig:fig2} display the cumulative returns for US and Japanese stocks, respectively, across the various portfolio strategies. In these figures, we label the HR--LR CPPS with AR($3$) as Conformal (AR), and the HR--LR CPPS with neural networks as Conformal (NN).

Figure~\ref{fig:fig1}, which shows the US stock results, indicates that Conformal (AR) and Conformal (NN) consistently outperform the other methods in terms of cumulative returns. Notably, these two approaches also exhibit better stability, avoiding the sharp drawdowns observed in methods such as Uniform and AR(3).

Figure~\ref{fig:fig2} presents results for the Japanese market and again confirms the strong performance of the Conformal approaches. Conformal (AR) achieves the highest cumulative returns, closely followed by Conformal (NN). Both methods effectively reduce the impact of extreme losses, as demonstrated by their smoother upward trajectories relative to portfolios such as Mean$_{t}[1]$ or AR(2), which display more pronounced fluctuations.

Overall, these findings underscore the robustness of the HR--LR CPPS framework in balancing risk and return. By incorporating predictive intervals to capture uncertainty, the proposed approach adapts to changing data distributions and maintains stable performance, particularly during market downturns. This adaptability stems from accounting for predictive uncertainty in constructing portfolios, in contrast to conventional methods (e.g., Mean$_{t}[3]$ and AR(1)) that rely primarily on point estimates.

It should be noted that the alternative methods considered here are relatively simple. Although more sophisticated approaches exist, they can introduce additional complexity, making direct comparisons less transparent. For our purposes, using straightforward baselines is appropriate.

\section{Conclusion}
In this study, we introduced a flexible framework for portfolio selection that employs conformal prediction  to generate prediction intervals. As a concrete illustration, we presented the HR--LR CPPS, which selects a portfolio exhibiting the highest potential return among those with favorable lower-bound risk profiles. Our empirical analyses with US and Japanese stock data suggest that HR--LR CPPS can effectively limit substantial drawdowns and maintain comparatively stable growth, underscoring the practical utility of conformal prediction  in navigating the inherent uncertainty of financial data. Overall, these findings highlight the value of predictive intervals in portfolio construction and support the broader applicability of our method to real-world investment scenarios.

\bibliographystyle{tmlr}
\bibliography{arXiv.bbl}

\onecolumn

\appendix

\end{document}